\documentclass[a4paper]{jpconf}
\usepackage{citesort}
\usepackage{graphicx}
\usepackage{amsmath} % align
\usepackage{rotating} % rotate
\begin{document}
\title{Lepton flavour violating stau decays versus seesaw parameters: correlations and expected number of events for both seesaw type-I and II}

%\author{Albert Villanova del Moral}
\author{A Villanova del Moral}

\address{Departamento de F\'\i sica and CFTP, Instituto Superior T\'ecnico, Avenida Rovisco Pais 1, 1049-001 Lisboa, Portugal}

\ead{albert@cftp.ist.utl.pt}

\begin{abstract}
In minimal supergravity (mSugra), the neutrino sector is related to the slepton sector by means of the renormalization group equations. This opens a door to indirectly test the neutrino sector via measurements at the LHC. Concretely, for the simplest seesaw type-I, we present the correlations between seesaw parameters and ratio of stau lepton flavour violating (LFV) branching ratios. We find some simple, extreme scenarios for the unknown right-handed parameters, where ratios of LFV rates correlate with neutrino oscillation parameters. On the other hand, we scan the mSugra parameter space, for both seesaw type-I and II, to find regions where LFV stau decays can be maximized, while respecting low-energy experimental bounds. We estimate the expected number of events at the LHC for a sample luminosity of ${\cal L} = 100 \mathrm{fb}^{-1}$.
\end{abstract}

%%%%%%%%%%%%%%%%%%%%%%%%%%%%%%%%%%%%%%%%%%%%%%%%%%%%%%%%%%%%%%%%%%%%%%%%%%%%%%%%%%%%%%%%%%%%%%%%%%%%%%%%%%%%%%%%%%%%%%%%%%%%%%%%%%%%%%%%%%%%%%%%%%%%%%%%%%%%%%%%%%%%%%
\section{Introduction}
%%%%%%%%%%%%%%%%%%%%%%%%%%%%%%%%%%%%%%%%%%%%%%%%%%%%%%%%%%%%%%%%%%%%%%%%%%%%%%%%%%%%%%%%%%%%%%%%%%%%%%%%%%%%%%%%%%%%%%%%%%%%%%%%%%%%%%%%%%%%%%%%%%%%%%%%%%%%%%%%%%%%%%
Many theoretical models aim at explaining what neutrino experiments~\cite{Fukuda:1998mi,:2008ee,Arpesella:2008mt,Aharmim:2008kc,Adamson:2008zt} have demonstrated: that neutrinos have small mass. Among the proposals to explain current neutrino data~\cite{Maltoni:2004ei,Schwetz:2008er}, the most popular ones are based in the seesaw mechanism ~\cite{Minkowski:1977sc,GellMann:1980vs,Yanagida:1979as,Mohapatra:1979ia,Schechter:1980gr,Schechter:1981cv}. Although this kind of models are not directly testable, they can be indirectly tested if we assume a supersymmetry (SUSY) framework with universal boundary conditions (like mSugra), as the Renormalization Group Equations (RGE's) relate the neutrino Yukawa couplings to the soft SUSY breaking slepton mass parameters.
Section~\ref{sec:correlations} shows the correlations between stau LFV branching ratios (BR's) and neutrino parameters. In the case of type-I seesaw, we only present the correlations obtained assuming degenerate heavy neutrinos. For a more complete analysis, see reference~\cite{Hirsch:2008dy}.
Similar correlations as the ones obtained in the simplest type-I SUSY seesaw can be obtained in a $SU(5)$ inspired type-II SUSY seesaw model~\cite{Rossi:2002zb}. For more details about the correlations, see reference~\cite{Hirsch:2008gh}.
Section~\ref{sec:scan} presents an analysis over the mSugra parameter space of the stau LFV decays and the estimated maximum number of events for the opposite-sign dilepton signal $\chi^0_2\to\chi^0_1\,\mu\,\tau$, for both type-I and II seesaw mechanisms.
%%%%%%%%%%%%%%%%%%%%%%%%%%%%%%%%%%%%%%%%%%%%%%%%%%%%%%%%%%%%%%%%%%%%%%%%%%%%%%%%%%%%%%%%%%%%%%%%%%%%%%%%%%%%%%%%%%%%%%%%%%%%%%%%%%%%%%%%%%%%%%%%%%%%%%%%%%%%%%%%%%%%%%
\section{Correlations for type-I seesaw}\label{sec:correlations}
%%%%%%%%%%%%%%%%%%%%%%%%%%%%%%%%%%%%%%%%%%%%%%%%%%%%%%%%%%%%%%%%%%%%%%%%%%%%%%%%%%%%%%%%%%%%%%%%%%%%%%%%%%%%%%%%%%%%%%%%%%%%%%%%%%%%%%%%%%%%%%%%%%%%%%%%%%%%%%%%%%%%%%
In section~\ref{sec:correlations-analytical} we will obtain analytical expressions for the correlations between neutrino parameters and stau LFV BR's, while in section~\ref{sec:correlations-numerical} we will check the validity of the analytical estimates via a full numerical calculation.
%%%%%%%%%%%%%%%%%%%%%%%%%%%%%%%%%%%%%%%%%%%%%%%%%%%%%%%%%%%%%%%%%%%%%%
%\subsection{Type-I seesaw}
%%%%%%%%%%%%%%%%%%%%%%%%%%%%%%%%%%%%%%%%%%%%%%%%%%%%%%%%%%%%%%%%%%%%%%
\subsection{Analytical correlations}\label{sec:correlations-analytical}
We briefly introduce the simplest type-I SUSY seesaw mechanism. 
Its particle content is the same as in the MSSM, but enlarged by three right-handed neutrino superfields ${\widehat N^c}_i$. The leptonic part of the superpotential is
\begin{equation}\label{su_pot}
W  =  Y_{e}^{ji}{\widehat L}_i{\widehat H_d}{\widehat E^c}_j
  + Y_{\nu}^{ji}{\widehat L}_i{\widehat H_u}{\widehat N^c}_j
  + M_{i}{\widehat N^c}_i{\widehat N^c}_i\,,
\end{equation}
where $Y_e$ and $Y_{\nu}$ denote the charged lepton and neutrino Yukawa couplings, while $M_{i}$ are the Majorana mass terms (of unspecified origin) of the ``right-handed'' neutrino superfields. From now on, we will work in the basis where both the Majorana mass matrix of the right-handed neutrinos and the charged lepton Yukawa coupling matrix are diagonal: $\hat M_R$  and $\hat Y_{e}$, respectively.

The effective mass matrix of the mostly left-handed neutrinos is given in the usual seesaw approximation as
\begin{equation}\label{meff}
m_{\nu} = - \frac{v_U^2}{2} Y_{\nu}^T\cdot \hat M_{R}^{-1}\cdot Y_{\nu}\,.
\end{equation}
This complex symmetrix matrix is diagonalized by the leptonic mixing matrix $U$~\cite{Schechter:1980gr}
\begin{equation}\label{diagmeff}
{\hat m_{\nu}} = U^T \cdot m_{\nu} \cdot U\,,
\end{equation}
where ${\hat m_{\nu}}=\textrm{diag}(m_i)$.
We can parametrize the neutrino Yukawa matrix as~\cite{Casas:2001sr}
\begin{equation}\label{Ynu}
Y_{\nu} =\sqrt{2}\frac{i}{v_U}\sqrt{\hat M_R}R\sqrt{{\hat m_{\nu}}}U^{\dagger},
\end{equation}
where $\hat m_{\nu}$ and $\hat M_R$ are diagonal matrices with the light neutrino mass eigenvalues $m_i$ and the heavy neutrino mass eigenvalues $M_i$, respectively; $U$ is the leptonic mixing matrix and $R$ is a complex orthogonal matrix.
 
The LFV branching ratios of the charged sleptons are related to the off-diagonal elements of their mass matrix. Neglecting left-right mixing in the charged slepton sector, these correspond to the off-diagonal elements of the slepton soft SUSY breaking mass parameter. In the general MSSM, the soft SUSY breaking parameters are free. However, if we assume an mSugra-like framework as the origin of SUSY breaking, soft SUSY breaking masses are diagonal at the GUT scale and they only get non-zero off-diagonal elements at the electroweak (EW) scale by means of the RGE's.
If the mixing between different flavour eigenstates of the charged sleptons is small (what we call the small-angle approximation), the LFV BR's for the decays of the charged sleptons can be approximated in terms of the off-diagonal soft breaking mass parameters%the square of the off-diagonal soft breaking mass-squared term
\begin{equation}
\textrm{BR}(\tilde l_i\to l_j\,\chi_1^0)\propto \left|(\Delta M_{\tilde L}^2)_{ij}\right|^2. %\propto \left|(Y_{\nu}^{\dagger}LY_{\nu})_{ij}\right|^2. 
\end{equation}
In the leading-log approximated solutions to the RGE's, the left-slepton LFV decays are proportional to
\begin{equation}
\textrm{BR}(\tilde l_i\to l_j\,\chi_1^0)%\propto \left|(\Delta M_{\tilde L}^2)_{ij}\right|^2 
\propto \left|(Y_{\nu}^{\dagger}LY_{\nu})_{ij}\right|^2, 
\end{equation}
where $L$ is defined as
\begin{equation}\label{deffacL}
L_{kl} = \log\Big(\frac{M_X}{M_{k}}\Big)\delta_{kl}
\end{equation}
and $M_X$ is the GUT scale.
Note that, in the leading-log approximation, the off-diagonal elements of the right-slepton soft breaking mass are zero, so that we will restrict our analysis to ``left-handed'' sleptons. Using the parametrization defined in equation~(\ref{Ynu}), the left-slepton LFV decays can be correlated to neutrino parameters
\begin{equation}
\textrm{BR}(\tilde l_i\to l_j\,\chi_1^0)\propto \left|U_{i\alpha}U_{j\beta}^*\sqrt{m_{\alpha}}\sqrt{m_{\beta}}
R_{k\alpha}^*R_{k\beta}M_k\log\left(\frac{M_X}{M_k}\right)\right|^2. 
\end{equation}
Taking \emph{ratios of LFV braching ratios}, we can eliminate the dependence on the SUSY parameters, so that we can establish clean correlations between ``left-handed'' slepton LFV BR's and neutrino parameters only. This way, for example, we can estimate that the ratio of the heaviest stau LFV decays is related to neutrino parameters as
\begin{equation}\label{eq:r1323-general}
\frac{Br({\tilde\tau}_2 \to  e \,\chi^0_1)}
     {Br({\tilde\tau}_2 \to  \mu \,\chi^0_1)}
 \simeq \left|r^{13}_{23}\right|^2 \equiv
\left|\frac{(Y_{\nu}^{\dagger}LY_{\nu})_{13}}{(Y_{\nu}^{\dagger}LY_{\nu})_{23}}\right|^2.
\end{equation}
We can use equation~(\ref{eq:r1323-general}) to obtain precise correlations between observables for different neutrino scenarios. If we assume that heavy neutrinos are degenerate, $R$ is real and that $U$ fulfills the tribimaximal (TBM) ansatz~\cite{Harrison:2002er}, then
\begin{equation}\label{eq:DegNuR-TBM}
r^{13}_{23}  = \frac{2 (m_2-m_1)}{|3 m_3-2 m_2-m_1|}.
\end{equation}
Table~\ref{tab:DegNuR-TBM} shows the form of equation~(\ref{eq:DegNuR-TBM}) and its numerical values, for different neutrino scenarios. %: strict normal hierrachy (SNH), strict inverse hierrachy (SIH) and quasidegenerate neutrinos (QD).
\begin{table}[htb]
\begin{center}
\caption{Parameter $r^{13}_{23}$ for the case of degenerate heavy neutrinos, assuming $R$ being real and TBM mixing for light neutrinos. Each column label corresponds to a different light neutrino scenario: strict normal hierarchical (SNH), strict inverse hierarchical (SIH), quasidegenerete normal hierarchical (QDNH) and quasidegenerete inverse hierarchical (QDIH) light neutrinos. Row label ANALYTICAL shows the analytical form of the parameter $r^{13}_{23}$; row label BFP shows its value when fixing neutrino mass splittings to their best fit point value~\cite{Maltoni:2004ei,Schwetz:2008er}; and row label $3\sigma$ shows its value when considering the $3\sigma$ allowed range for the neutrino mass-squared differences. Note that $\alpha \equiv \Delta m^2_{\textsc{s}}/|\Delta m^2_{\textsc{a}}|$ is the ratio of the solar over the atmospheric mass splitting and $\sigma_{\textsc{a}}\equiv\Delta m^2_{\textsc{a}}/|\Delta m^2_{\textsc{a}}|$ is the sign of the atmospheric mass splitting.}
\begin{tabular}{lll}\br%\cline{2-3}%\hline
%{\small
%\multicolumn{1}{c}{} 
& {\bf SNH} & {\bf SIH} \\\mr%\cline{2-3}%\hline
%
%\multicolumn{1}{c}{} 
{\bf ANALYTICAL}& $r^{13}_{23} = \frac{2\sqrt{\alpha}}{3\sqrt{1+\alpha}-2\sqrt{\alpha}}$ & 
$r^{13}_{23} = \frac{2(1-\sqrt{1-\alpha})}{2+\sqrt{1-\alpha}}$ \\%\hline
{\bf BFP} & $(r^{13}_{23})^2 = 1.7\times 10^{-2}$ & $(r^{13}_{23})^2 = 1.1\times 10^{-4}$ \\%\hline
$\boldsymbol{3\sigma}$ & $(r^{13}_{23})^2 \in [0.92,\,3.7]\times 10^{-2}$ & $(r^{13}_{23})^2 \in [0.48,\,3.3]\times 10^{-4}$ \\\br%\hline
%}
& {\bf QDNH} & {\bf QDIH} \\\mr%\cline{2-3}%\hline
{\bf ANALYTICAL} & $r^{13}_{23} \simeq \frac{2\alpha}{3\sigma_{\textsc{a}}+\alpha}$ & $r^{13}_{23} \simeq \frac{2\alpha}{3\sigma_{\textsc{a}}+\alpha}$ \\%\hline
{\bf BFP} & $(r^{13}_{23})^2 = 4.4\times 10^{-4}$ & $(r^{13}_{23})^2 = 4.6\times 10^{-4}$ \\%\hline
$\boldsymbol{3\sigma}$ & $(r^{13}_{23})^2 \in [1.8,\,12]\times 10^{-4}$ & $(r^{13}_{23})^2 \in [1.9,\,13]\times 10^{-4}$ \\\br
\end{tabular}
\label{tab:DegNuR-TBM}
\end{center}
\end{table}
For a general analysis of different neutrino hierarchies, dependence on the overall light neutrino mass scale, departure from tribimaximality, both for degenerate and strongly hierarchical heavy neutrinos, see reference~\cite{Hirsch:2008dy}.
%%%%%%%%%%%%%%%%%%%%%%%%%%%%%%%%%%%%%%%
\subsection{Numerical correlations}\label{sec:correlations-numerical}
In order to check the analytical correlations obtained in section~\ref{sec:correlations-analytical}, we have performed a numerical calculation using the program package \textsc{SPheno}~\cite{Porod:2003um}. Although we have checked that the validity of the analytical correlations is independent of the mSugra parameters, we only present the results for some specific benchmark points. A more datailed study over the mSugra parameter space is presented in section~\ref{sec:scan}.
The numerical procedure followed to fit neutrino data consists in first fixing the neutrino Yukawa couplings at the GUT scale by means of equation~(\ref{Ynu}) and assuming certain neutrino scenarios as input parameters. The RGE's are numerically run with \textsc{SPheno}, so that the light neutrino masses are calculated at the EW scale as output parameters. An iterative process varies the input parameters at the GUT scale in order to obtain the desired output parameters at the EW scale.

\begin{figure}[htbp]
\begin{center}
\vspace{5mm}
\includegraphics[width=0.48\textwidth]{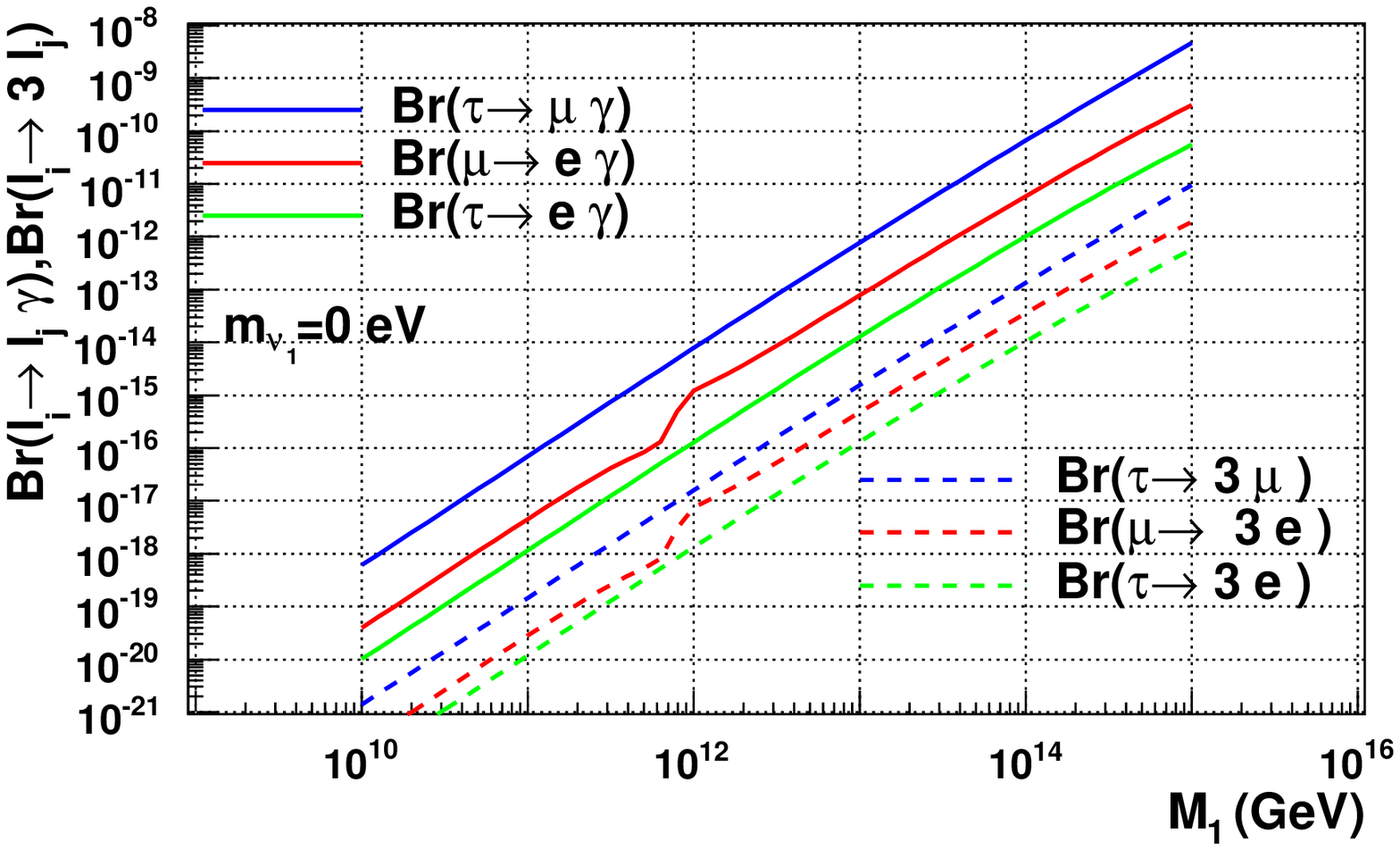}
\includegraphics[width=0.48\textwidth]{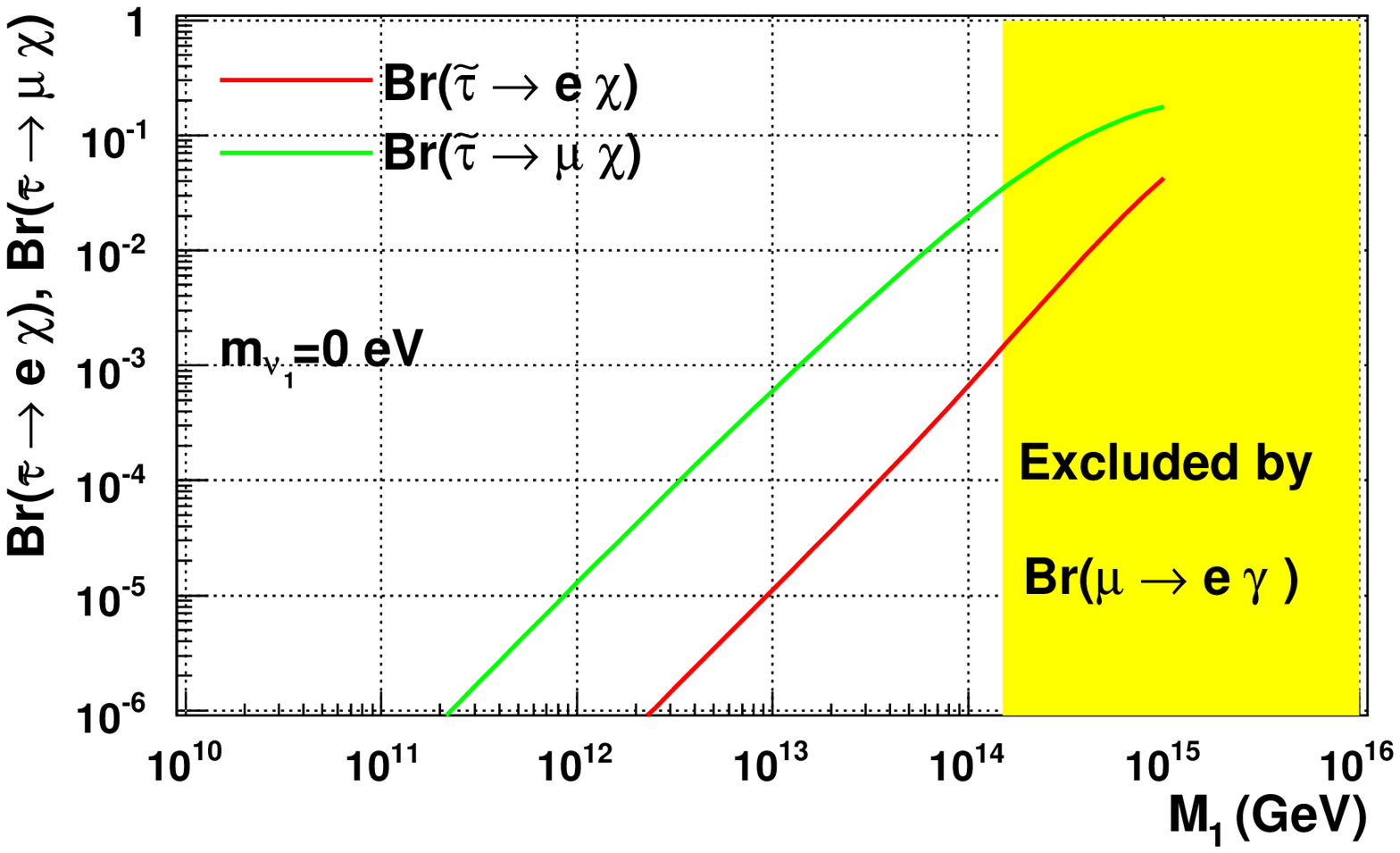}
\end{center}
\caption{Branching ratios for $l_i \to l_j \, \gamma$ and $l_i \to 3 l_j$ (left), and ${\tilde\tau}_2 \to e \,\chi^0_1$ and ${\tilde\tau}_2\to \mu \,\chi^0_1$ (right) for the standard point SPS3 versus $M_R$, assuming degenerate right-handed neutrinos. Neutrino oscillation parameters have been fixed to the best fit values for the mass splittings~\cite{Maltoni:2004ei,Schwetz:2008er}, with exact TBM neutrino angles. The absolute neutrino mass scale has been fixed to $m_1=0$.  The coloured region in the right-side plot is excluded from the current experimental limit on BR($\mu\to e\,\gamma$).}
\label{fig:BrSPS3}
\end{figure}

Figure~\ref{fig:BrSPS3} shows the dependence of LFV processes (low energy LFV processes in the left panel and stau LFV decays in the right panel) on the common mass $M_R$ of the degenerate heavy neutrinos, for the mSugra benchmark point SPS3~\cite{Allanach:2002nj}. 
We have checked that our analytical estimate of ratios of stau LFV branching ratios is in very good agreement with the numerically calculated ratio, as long as the magnitude of the stau LFV BR's are not larger than 10\% (where the small-angle approximation is no longer valid). 
It is important to note that current experimental upper bounds on low energy LFV processes put a severe constraint in the maximum possible value of stau LFV decays. Given the current experimental limit on BR($\mu\to e \,\gamma$)~\cite{Amsler:2008zzb}
\begin{equation}\label{eq:BRMuToEGammaBound}
\textrm{BR}(\mu\to e \gamma) \le 1.2\cdot 10^{-11}\,,
\end{equation}
the maximum possible value for  BR(${\tilde\tau}_2\to \mu \,\chi^0_1$) is a few percent for SPS3.

On the contrary, for benchmark point SPS1a'~\cite{AguilarSaavedra:2005pw}, the allowed maximum value for BR(${\tilde\tau}_2\to \mu\, \chi^0_1$) is only $10^{-3}$. 
The reason for this difference between points SPS3 and SPS1a' is that the SUSY mass spectrum is nearly two times heavier in the former. As BR($\mu\to e\, \gamma$) depends inversely on the SUSY spectrum~\cite{Hisano:1995cp}, %$\textrm{BR}(\mu \to e\,\gamma)\propto 1/ m_{\textrm{SUSY}}^8$, 
\begin{equation}
\textrm{BR}(\mu \to e\,\gamma)\propto \frac{1}{m_{\textrm{SUSY}}^8 }\,,
\end{equation}
its value is several hundred times smaller for SPS3, thus providing a weaker constraint on ${\tilde\tau}_2$ LFV BR's.

\begin{figure}
\begin{center}
\includegraphics[width=0.48\textwidth]{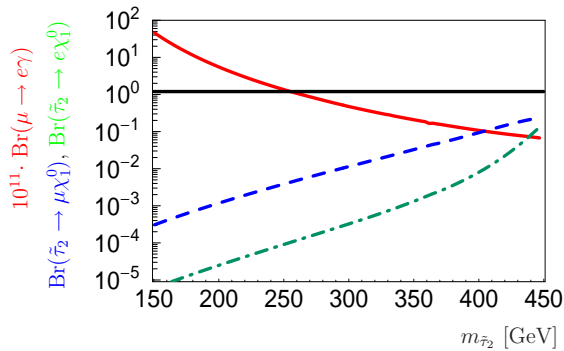}%height=60mm
\end{center}
\vskip1mm
\caption{LFV BR's obtained as function of the heavier scalar tau mass. The full line (red) is $10^{11} \times \textrm{BR}(\mu\to e\,\gamma)$, the dashed line (blue) BR(${\tilde\tau_2}\to\mu\,\chi^0_1$) and the dot-dashed line (green) is BR(${\tilde\tau_2}\to e\,\chi^0_1$).  Data obtained for SPS1a with parameters varied along the corresponding ``slope''. Right-handed neutrino mass is fixed to $M_R =  3\times 10^{13}$ GeV. Neutrino oscillation parameters have been fixed to the best fit values for the mass splittings~\cite{Maltoni:2004ei,Schwetz:2008er}, with exact TBM neutrino angles, and the absolute neutrino mass scale has been fixed to $m_1=0$. The black line is the current upper limit on BR($\mu \to  e\gamma$).}
\label{fig:Brs}
\end{figure}

The dependence of BR($\mu \to e \,\gamma$) on the SUSY spectrum is shown in figure~\ref{fig:Brs}. 
Here we plot BR($\mu \to e\,\gamma$), BR(${\tilde\tau_2}\to\mu\,\chi^0_1$) and
BR(${\tilde\tau_2} \to e\, \chi^0_1$) versus the mass of
${\tilde\tau}_2$. The range of $m_{{\tilde\tau}_2}$ has been obtained by 
varying the SPS1a parameters along the ``slope''~\cite{Allanach:2002nj}. 
Note that BR($\mu \to e\gamma$) decreases as the SUSY spectrum gets heavier, dropping
below its current experimental bound for $m_{{\tilde\tau}_2}$ larger
than about $250$ GeV. 
In contrast, ${\tilde\tau}_2$ LFV BR's increase for increasing $m_{{\tilde\tau}_2}$, as charged left-sleptons become increasingly degenerate and thus their mixing increases. 
Nevertheless, note also that the ratio of ${\tilde\tau}_2$ LVF BR's remains constant, in agreement with the analytical estimate, as long as the magnitude of each individual BR is not too large (and hence, the small-angle approximation is valid).
%%%%%%%%%%%%%%%%%%%%%%%%%%%%%%%%%%%%%%%%%%%%%%%%%%%%%%%%%%%%%%%%%%%
%\section{Correlations for type-II seesaw}
%%%%%%%%%%%%%%%%%%%%%%%%%%%%%%%%%%%%%%%%%%%%%%%%%%%%%%%%%%%%%%%%%%%
%Similar correlations as the ones obtained in the simplest SUSY seesaw type-I can be obtained in a $SU(5)$ inspired SUSY seesaw type-II model~\cite{Rossi:2002zb}. For more details see reference~\cite{Hirsch:2008gh}.
%%%%%%%%%%%%%%%%%%%%%%%%%%%%%%%%%%%%%%%%%%%%%%%%%%%%%%%%%%%%%%%%%%%%%%%%%%%%%%%%%%%%%%%%%%%%%%%%%%%%%%%%%%%%%%%%%%%%%%%%%%%%%%%%%%%%%%%%%%%%%%%%%%%%%%%%%%%%%%%%%%%%%%
\section{Scan}\label{sec:scan}
%%%%%%%%%%%%%%%%%%%%%%%%%%%%%%%%%%%%%%%%%%%%%%%%%%%%%%%%%%%%%%%%%%%%%%%%%%%%%%%%%%%%%%%%%%%%%%%%%%%%%%%%%%%%%%%%%%%%%%%%%%%%%%%%%%%%%%%%%%%%%%%%%%%%%%%%%%%%%%%%%%%%%%
In section~\ref{sec:correlations} it has been shown that it is possible to indirectly test the seesaw mechanism at the LHC by means of correlations between the ratio of ${\tilde\tau}_2$ LFV BR's and neutrino parameters. Although the magnitudes of ${\tilde\tau}_2$ LFV BR's have been studied for two specific SUSY benchmark points (SPS1a' and SPS3), a more general study over the mSugra parameter space is necessary. Thus, we can identify regions of the mSugra parameter space that maximize the magnitude of ${\tilde\tau}_2$ LFV BR's. Besides this, we have estimated the maximum number of events of the opposite-sign dilepton signal $\chi^0_2\to\chi^0_1\,\mu\,\tau$, which can be searched for at the LHC. Note that a complete Monte Carlo analysis would be needed, but this is out of the scope of this work.
The numerical procedure consists in scanning over the $m_0$-$m_{1/2}$ plane, for fixed values of other mSUGRA parameters, using the program package \textsc{SPheno}~\cite{Porod:2003um}. For each point in this plane, we perform a maximization of BR($\mu\to e \gamma$), so that it is as close as possible to its current experimental upper bound, given in equation (\ref{eq:BRMuToEGammaBound}). To fit neutrino data we have followed the same iterative procedure described in section~\ref{sec:correlations-numerical}. In our analysis we have always fitted light neutrino mass splittings to their best fit point values~\cite{Maltoni:2004ei,Schwetz:2008er}, under a strictly normal hierarchy ($m_1=0$) and their mixing to be TBM.
To calculate the number of events of the opposite-sign dilepton signal $\chi^0_2\to\chi^0_1\,\mu\,\tau$, we have estimated the total production cross section of $\chi^0_2$ at leading order with the package \textsc{Prospino}~\cite{Beenakker:1996ch,Beenakker:1997ut,Beenakker:1999xh,Spira:2002rd,Plehn:2004rp}.
%%%%%%%%%%%%%%%%%%%%%%%%%%%%%%%%%%%%%%%%%%%%%%%%%%%%%%%%%%%%%%%%%%%
\subsection{Type-I seesaw}\label{sec:scan-I}
\begin{figure}[htb]
\centering
\begin{tabular}{|c|c|c|}\cline{2-3}%\hline%\vspace{0.1cm}
%\hspace{0.6cm}\B 
\multicolumn{1}{c|}{}& \vphantom{\Large Ap}{\bf BR(${\tilde\tau}_2 \to \mu\,\chi^0_1$)} & {\bf BR(${\tilde\tau}_2 \to e\,\chi^0_1$)} \\\hline
\hspace{0.15cm}
\begin{rotate}{90}{{\small$\quad\ \tan\beta=10,\quad  A_0=0,\quad  \mu>0$}}\end{rotate} &
\includegraphics[width=0.35\textwidth]{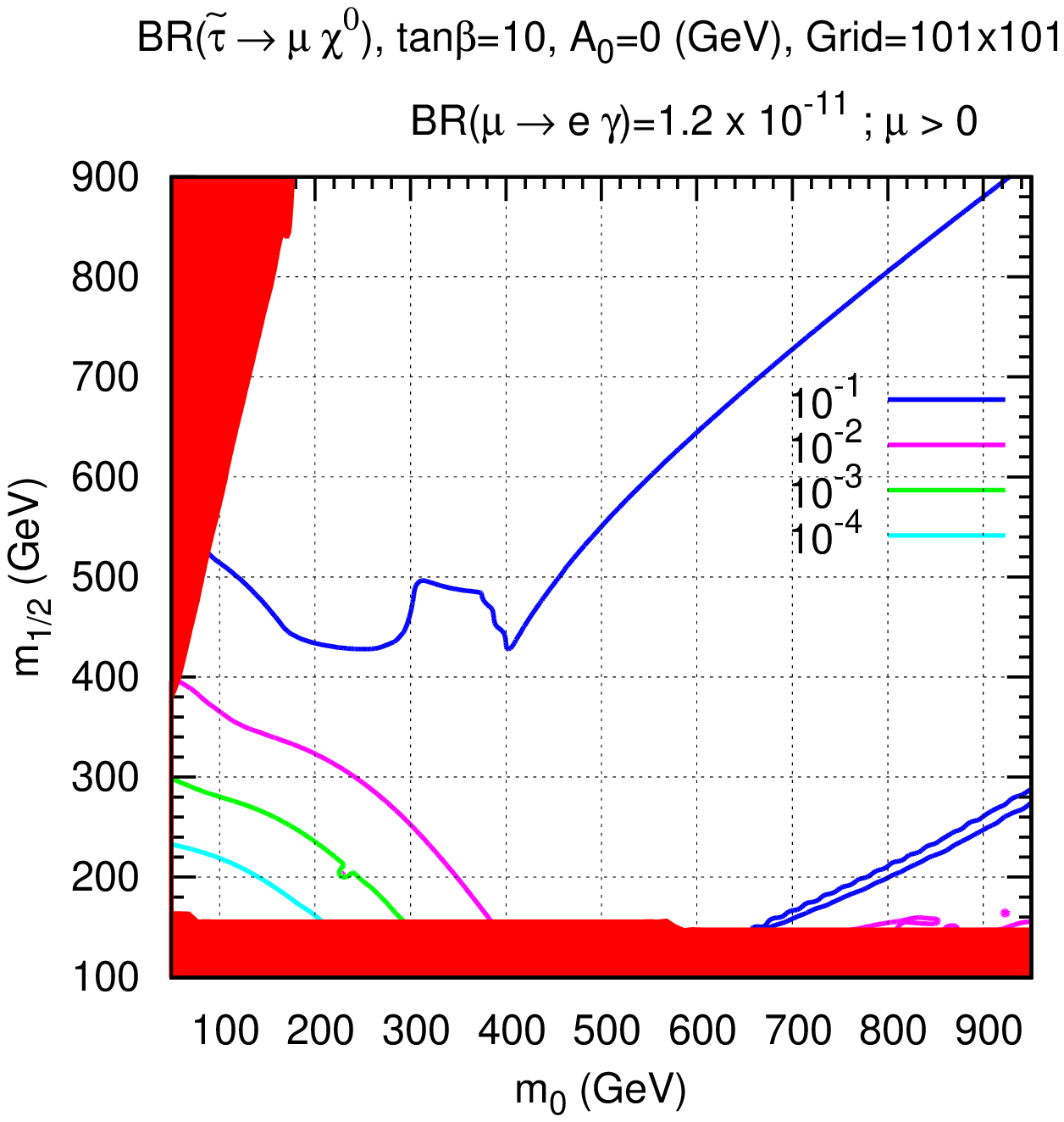}&
\includegraphics[width=0.35\textwidth]{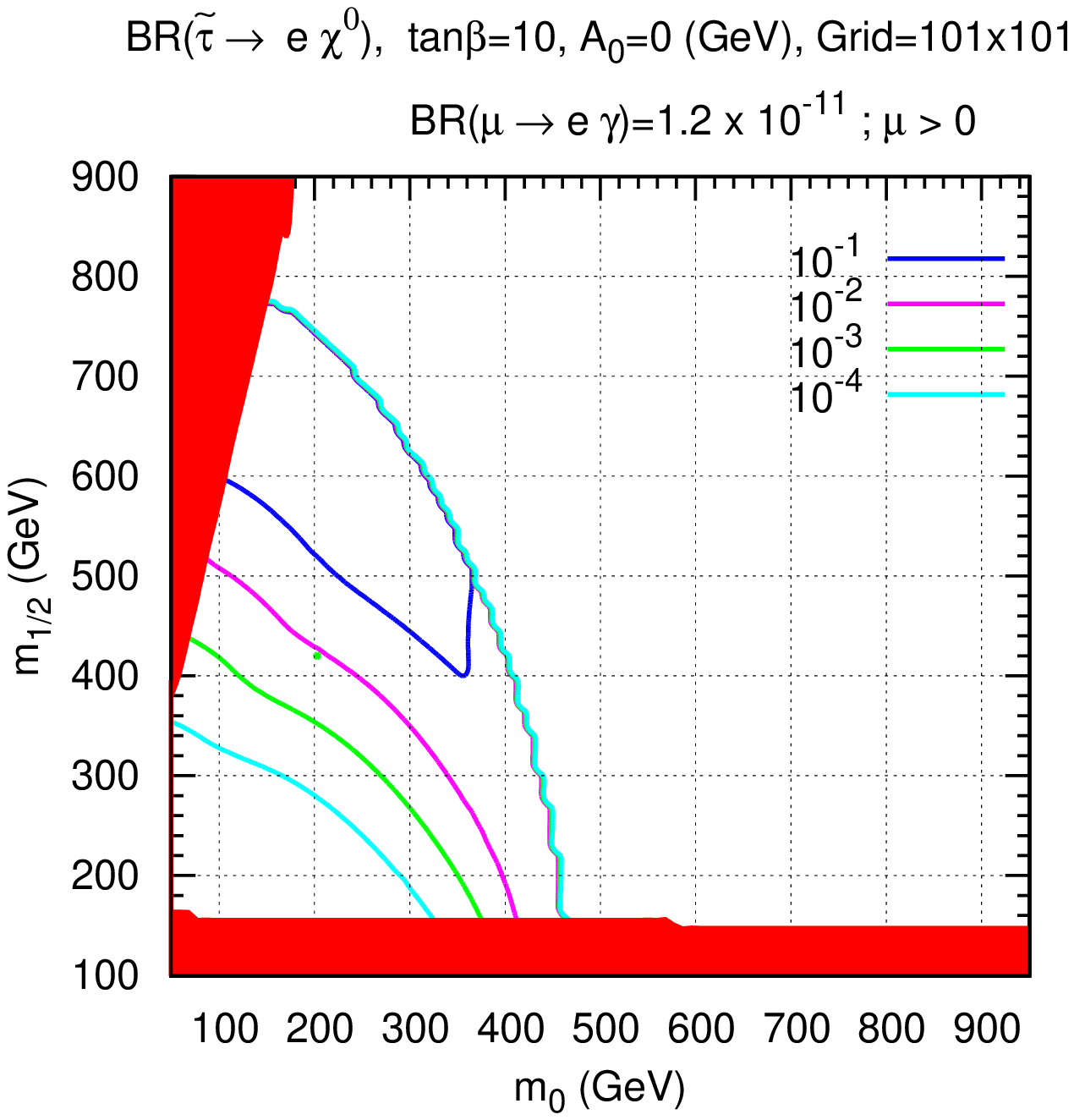}\\\hline
\hspace{0.30cm}%$\vspace{-2cm}
\begin{rotate}{90}{{\small$\quad\ \tan\beta=30,\quad  A_0=0,\quad  \mu>0$}}\end{rotate} &
  \includegraphics[width=0.35\textwidth]{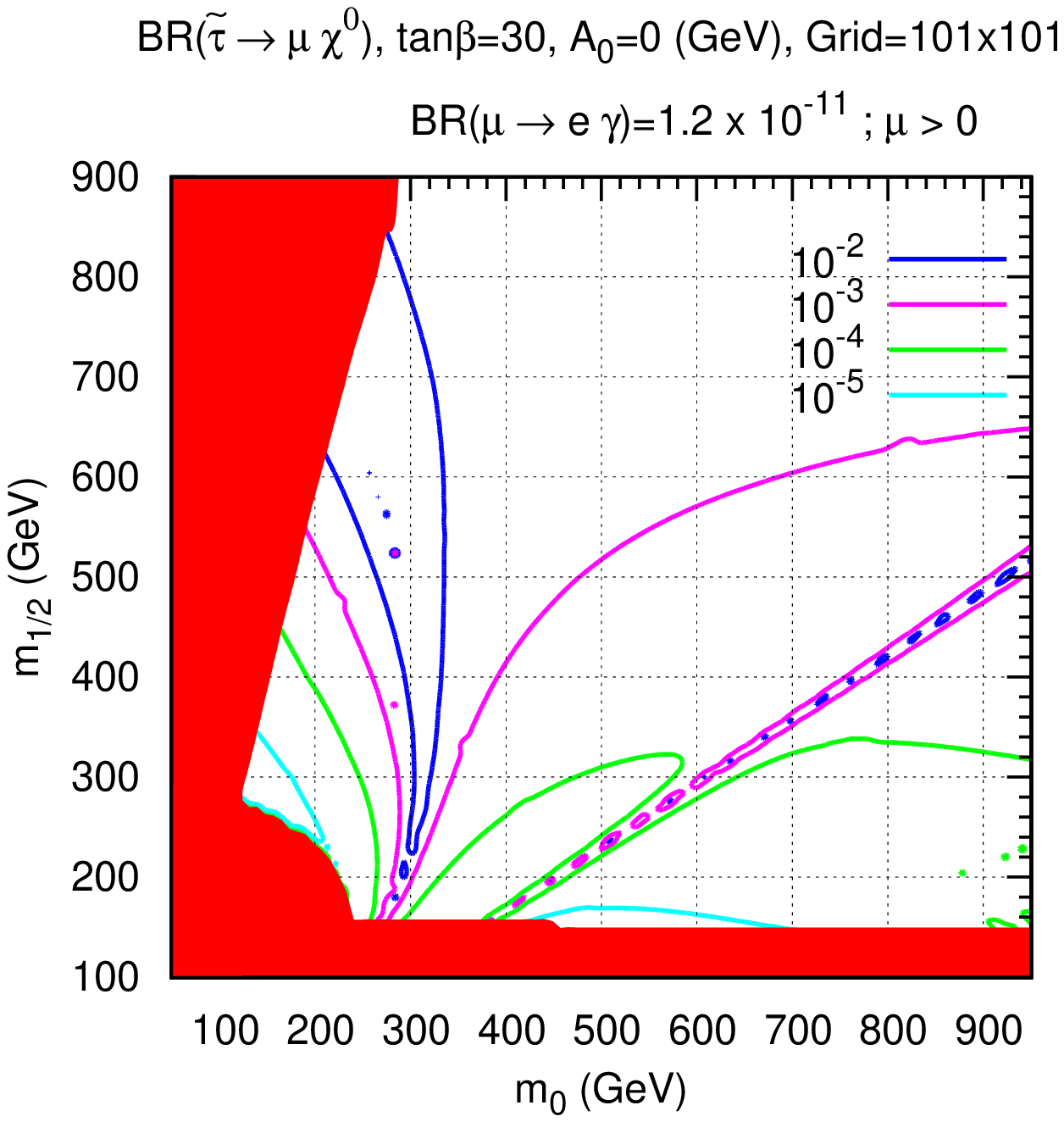} &
  \includegraphics[width=0.35\textwidth]{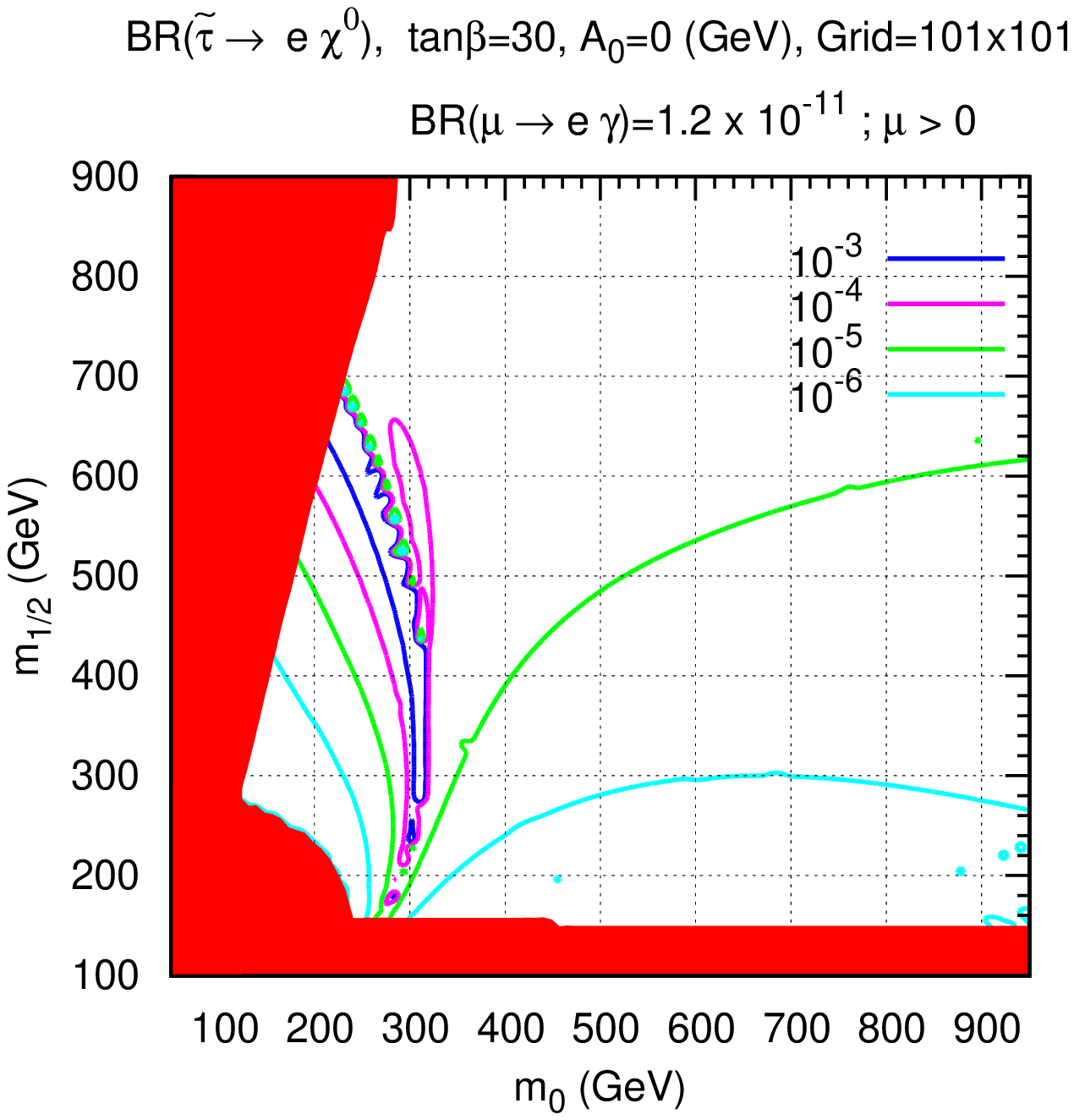}\\\hline
\hspace{0.30cm}%$\vspace{-2cm}
\begin{rotate}{90}{{\small$ \tan\beta=10,\  A_0=-300\textrm{ GeV},\  \mu>0$}}\end{rotate} &
  \includegraphics[width=0.35\textwidth]{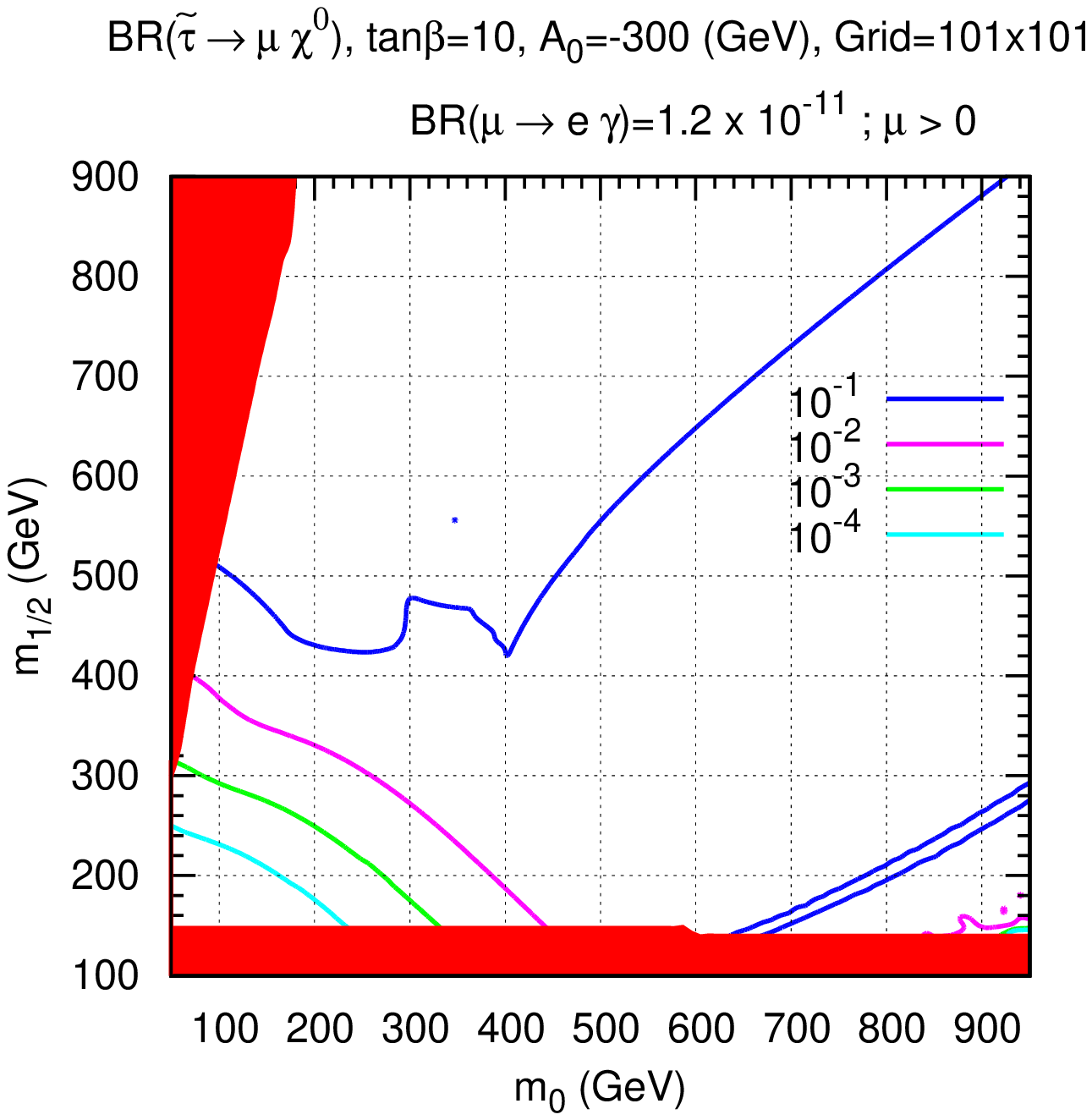}&
  \includegraphics[width=0.35\textwidth]{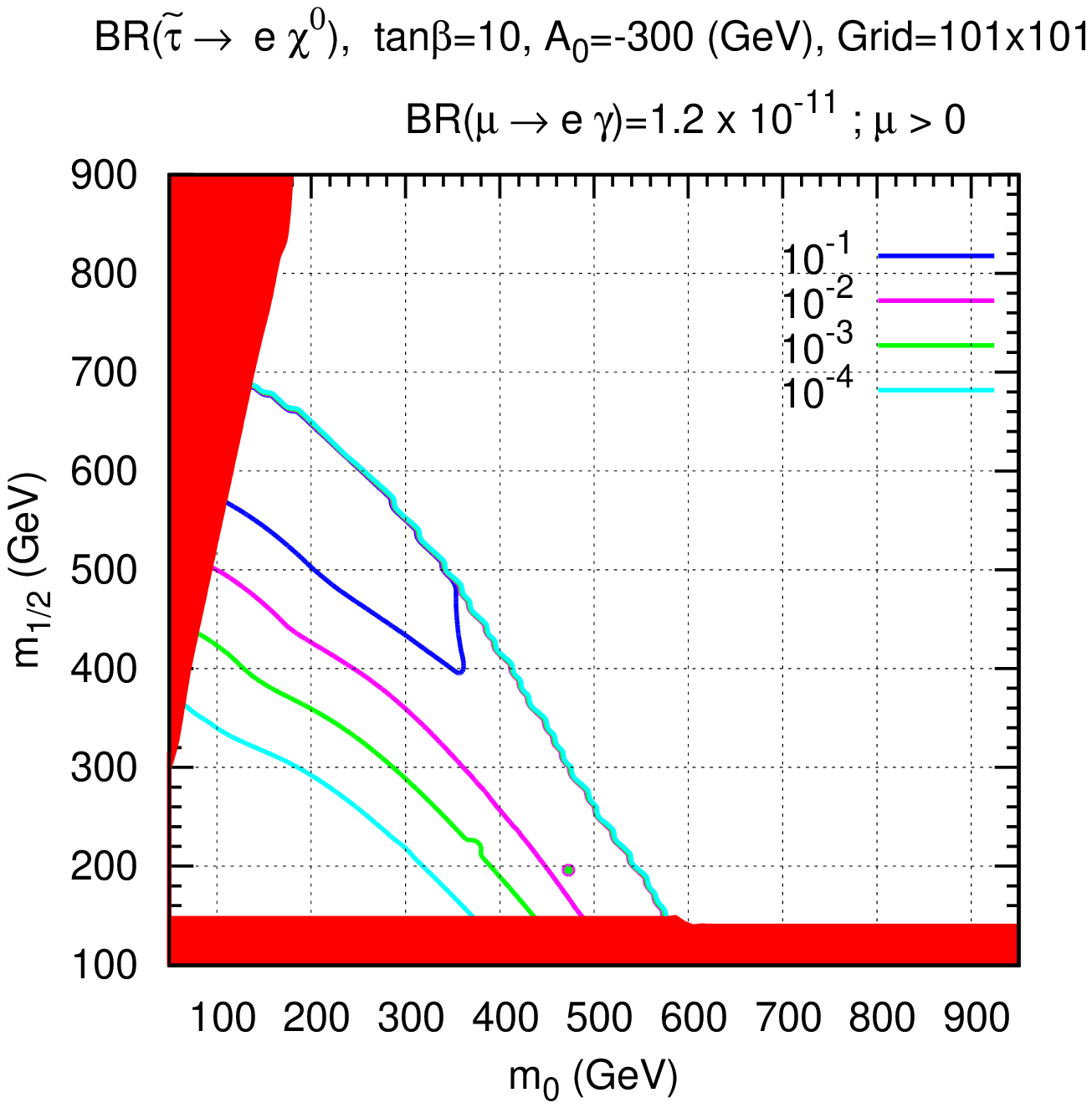}\\\hline
\end{tabular}
  \caption{BR(${\tilde\tau}_2 \to \mu\,\chi^0_1$) (left column) and
    BR(${\tilde\tau}_2 \to e\,\chi^0_1$) (right column) in the $m_0-m_{1/2}$ plane, 
    for the type-I seesaw. We impose BR($\mu\to e\, \gamma) \le 1.2\cdot
    10^{-11}$, and we consider different choice of mSugra parameters:
     $\mu>0$, $\tan\beta=10$ and $A_0=0$ GeV (first row);
     $\mu>0$, $\tan\beta=30$ and $A_0=0$ GeV (second row);
     $\mu>0$, $\tan\beta=10$ and $A_0=-300$ GeV (third row).
    } 
  \label{fig:StauLFV-table-I}
\end{figure}

%%%%%%%%%%%%%%%%%%%%%%%%%%%%%%%%%%%%%%%%%%%%%%%%%%%%%%%%%%%%%%%%%%%
In order to simplify our numerical analysis in type-I seesaw, we have only considered a specific neutrino scenario, assuming that right-handed neutrinos are degenerate and the matrix $R$ is real.
Figure~\ref{fig:StauLFV-table-I} shows the magnitude of BR(${\tilde\tau}_2 \to \mu\,\chi^0_1$) (left column) and BR(${\tilde\tau}_2 \to e\,\chi^0_1$) (right column) in the $m_0-m_{1/2}$ plane, for three combinations of mSugra parameters: the first row corresponds to what we call our standard mSugra point, defined by
\begin{equation}\label{eq:standard-msugra-point}
\tan\beta=10,\qquad  A_0=0,\qquad \mu>0\,;
\end{equation}
the second row corresponds to the same standard mSugra point, but with $\tan\beta=30$; the third row corresponds to the same standard mSugra point, but with $A_0=-300$ GeV. 
One important feature is that there is a region in the $m_0-m_{1/2}$ plane (generally, small $m_0$ and large $m_{1/2}$ values) where ${\tilde\tau}_2$ LFV BR's can be as large as 10\%. This is the parameter space region where the constraint imposed by BR($\mu\to e\,\gamma$) is supressed due to a strong cancellation between its neutralino-charged slepton and the chargino-sneutrino amplitude contributions. 
Note that, as previously shown in section~\ref{sec:correlations-analytical}, the BR(${\tilde\tau}_2 \to e\,\chi^0_1$) is normally smaller than the BR(${\tilde\tau}_2 \to \mu\,\chi^0_1$) by a factor $1.7\times 10^{-2}$. 
While ${\tilde\tau}_2$ LFV BR's are decreased by larger values of $\tan\beta$, they are practically not affected by $A_0$.
We have checked that the ${\tilde\tau}_2$ LFV BR's are not sensitive to the sign of neither $A_0$ nor $\mu$.

\begin{figure}[htb]
  \centering
  \includegraphics[width=0.75\textwidth]{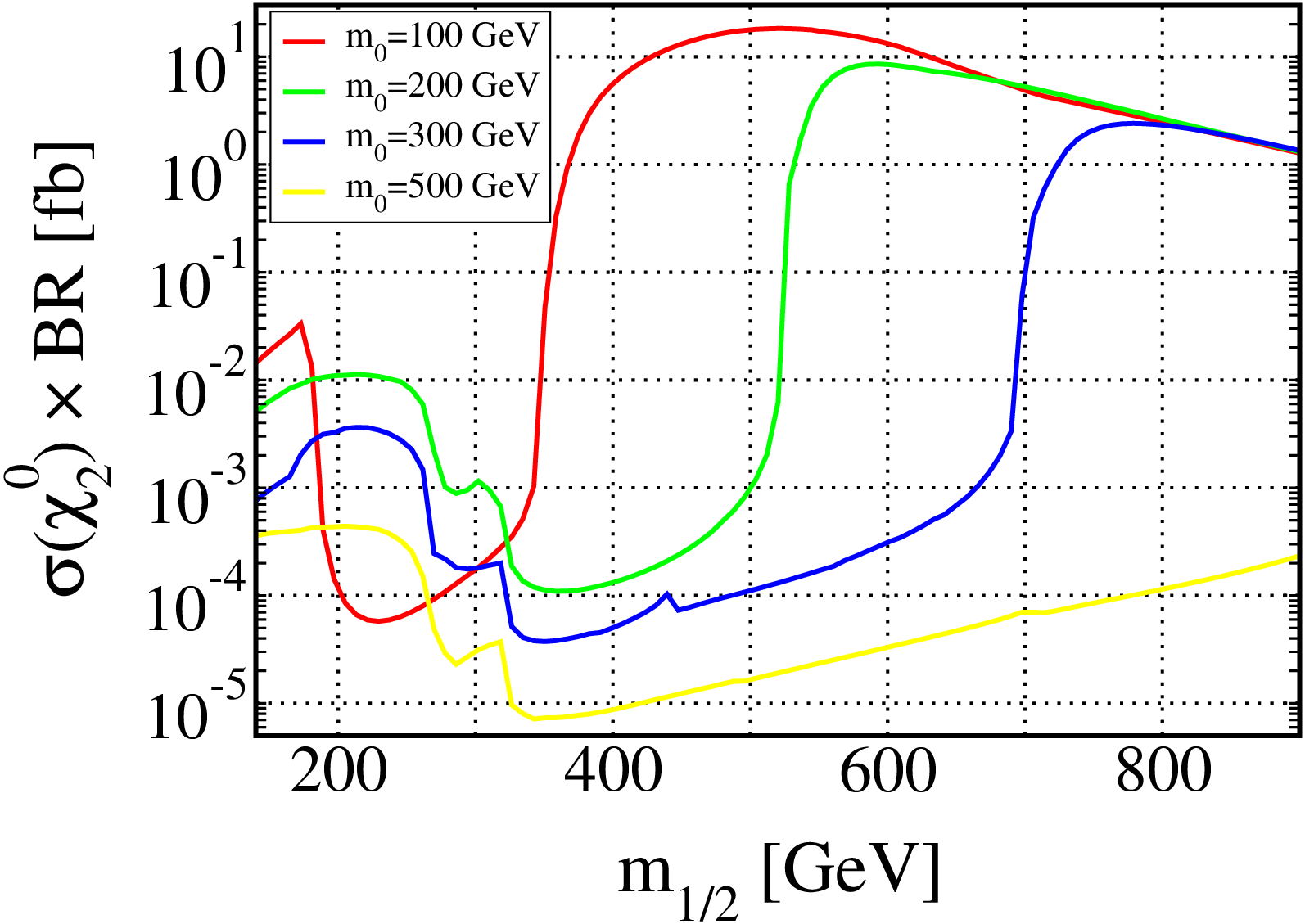}
  \caption{Production cross section (at leading order) of $\chi^0_2$ times BR($\chi^0_2\to\chi^0_1\,\mu\,\tau$)
    %of $\chi^0_2$ going to $\mu$-$\tau$ opposite-sign dilepton signal 
    versus $m_{1/2}$ for $m_0=100$~GeV (red),
    200~GeV (green), 300~GeV (blue) and 500~GeV (yellow). We take a standard choice of
    parameters: $\mu>0$, $\tan\beta=10$ and $A_0=0$ GeV, for type-I seesaw, for
    BR($\mu\to e \gamma) \le 1.2\cdot 10^{-11}$.} 
  \label{fig:ProdXBR-I}
\end{figure}

In figure~\ref{fig:ProdXBR-I}, we have plotted the production cross section $\sigma(\chi^0_2)$ at leading order times the BR of $\chi^0_2$ going to the opposite-sign dilepton signal $\chi^0_1\,\mu\,\tau$ as a function of $m_{1/2}$, for different values of $m_0$. We have fixed the rest of the mSugra parameters to our standard point defined in equation~(\ref{eq:standard-msugra-point}).
For $m_0\sim 100$ GeV and $m_{1/2}\sim[450,\,600]$ GeV and assuming a luminosity ${\cal L} = 100 fb^{-1}$, the number of events of the opposite-sign dilepton signal $\chi^0_2\to\chi^0_1\,\mu\,\tau$ can be of the order of $10^3$.
%%%%%%%%%%%%%%%%%%%%%%%%%%%%%%%%%%%%%%%%%%%%%%%%%%%%%%%%%%%%%%%%%%%%%%
\subsection{Type-II seesaw}\label{sec:scan-II}
\begin{figure}[htb]
\centering
\begin{tabular}{|c|c|c|}\cline{2-3}%\hline%\vspace{0.1cm}
%\hspace{0.6cm}\B 
\multicolumn{1}{c|}{}& \vphantom{\Large Ap}{\bf BR(${\tilde\tau}_2 \to \mu \chi^0_1$)} & {\bf BR(${\tilde\tau}_2 \to e \chi^0_1$)} \\\hline
\hspace{0.15cm}
\begin{rotate}{90}{{\small$\quad\ \tan\beta=10,\quad  A_0=0,\quad  \mu>0$}}\end{rotate} &
\includegraphics[width=0.35\textwidth]{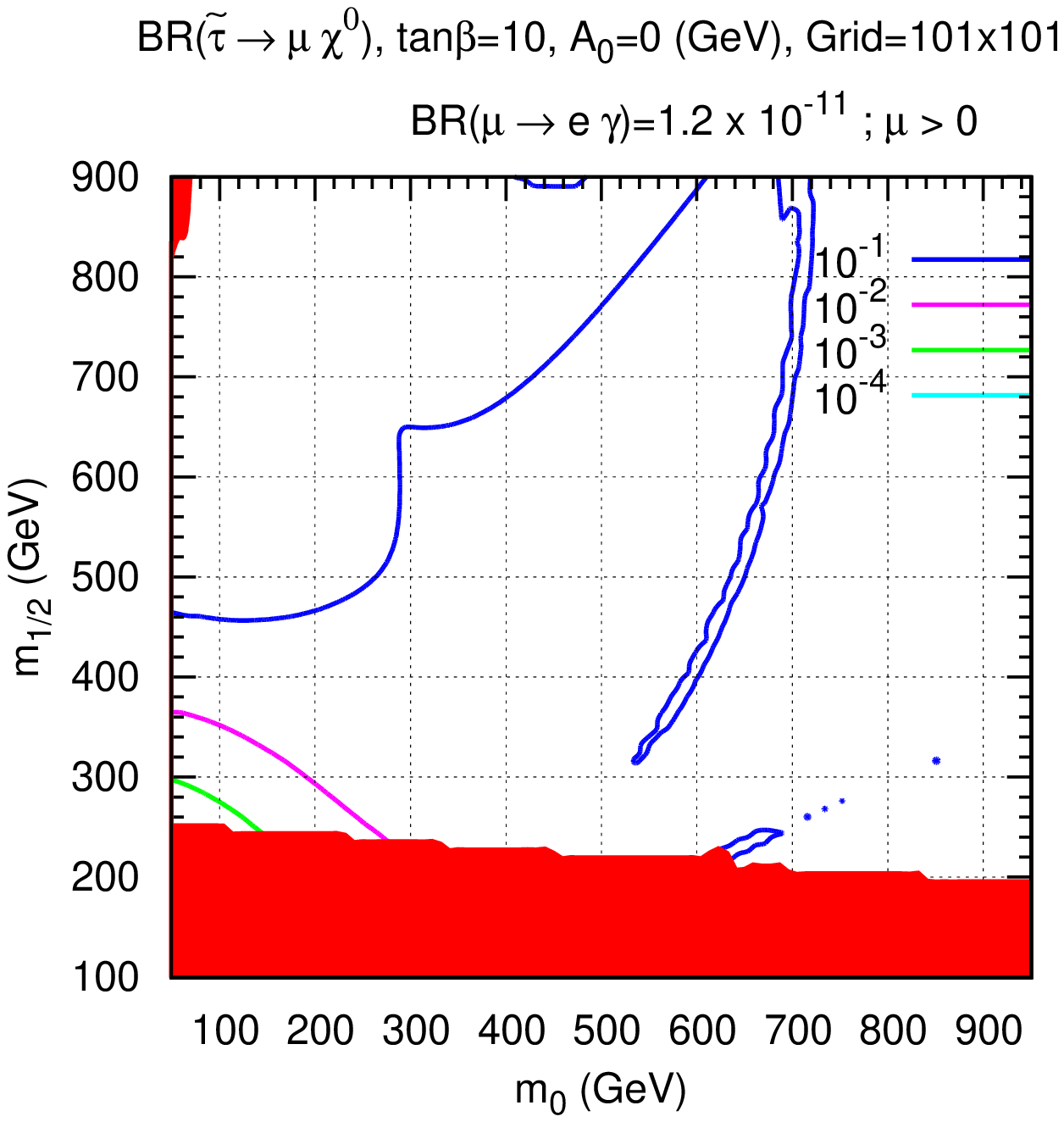}&
\includegraphics[width=0.35\textwidth]{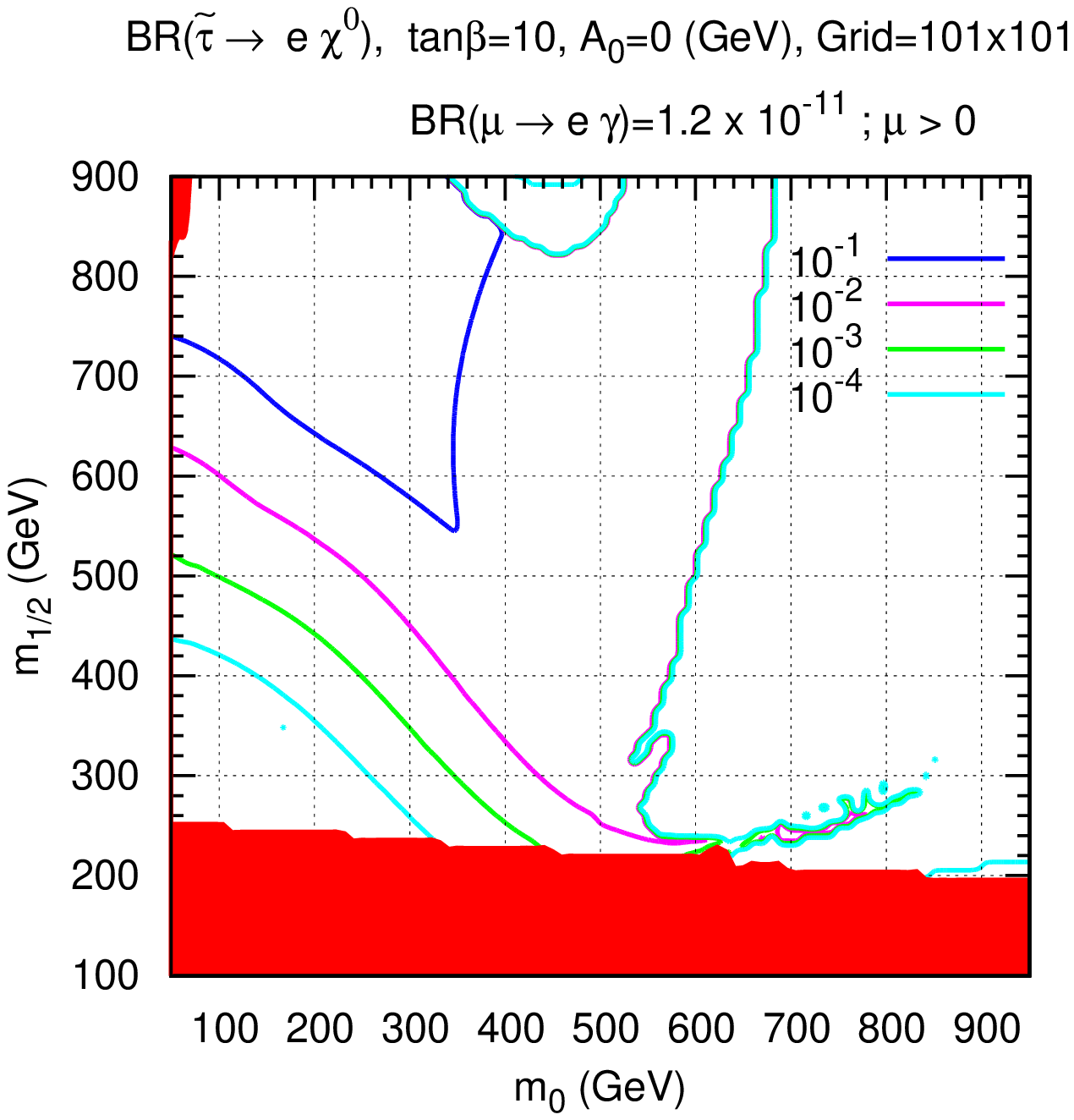}\\\hline
\hspace{0.30cm}%$\vspace{-2cm}
\begin{rotate}{90}{{\small$\quad\ \tan\beta=30,\quad  A_0=0,\quad  \mu>0$}}\end{rotate} &
  \includegraphics[width=0.35\textwidth]{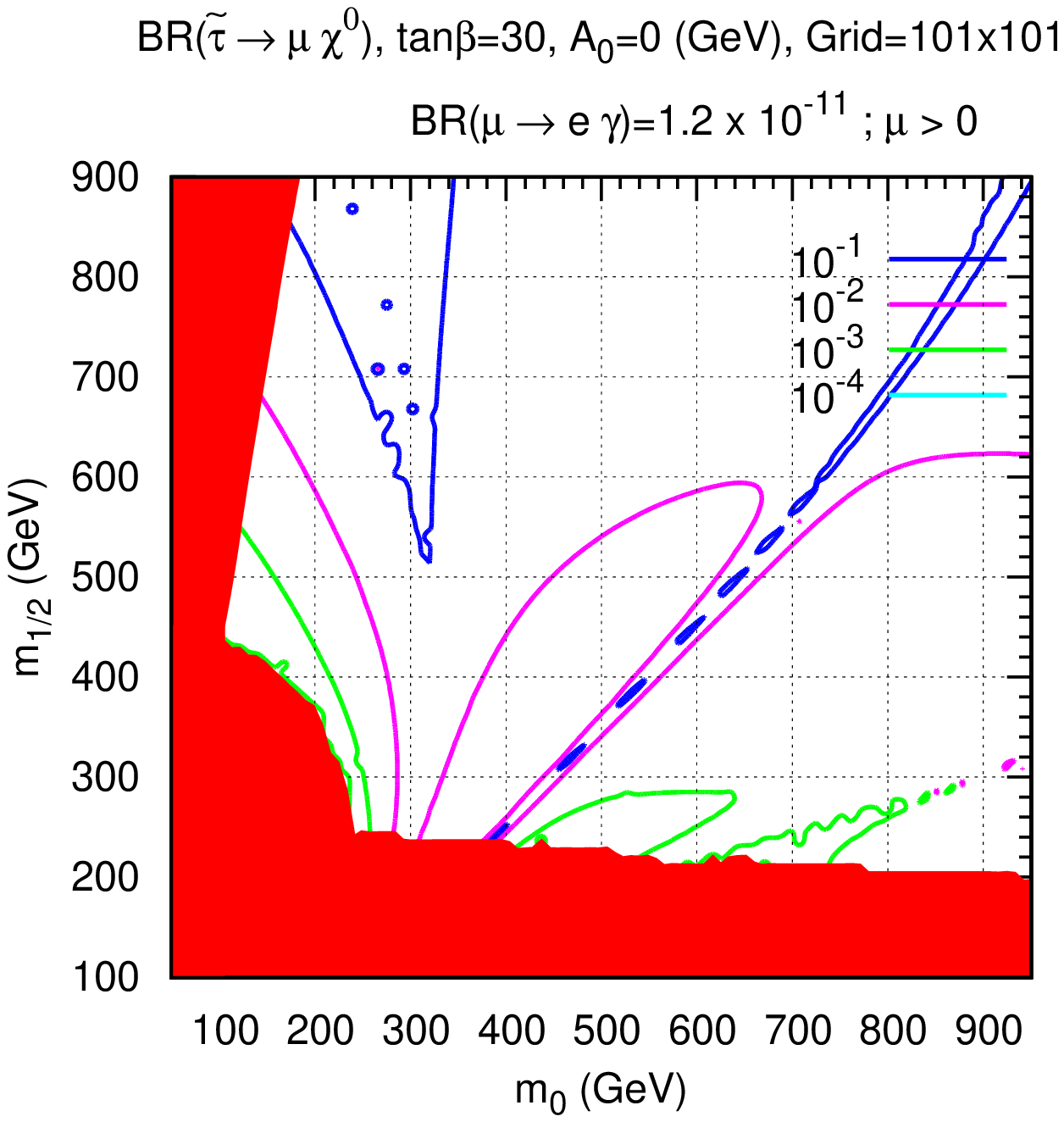} &
  \includegraphics[width=0.35\textwidth]{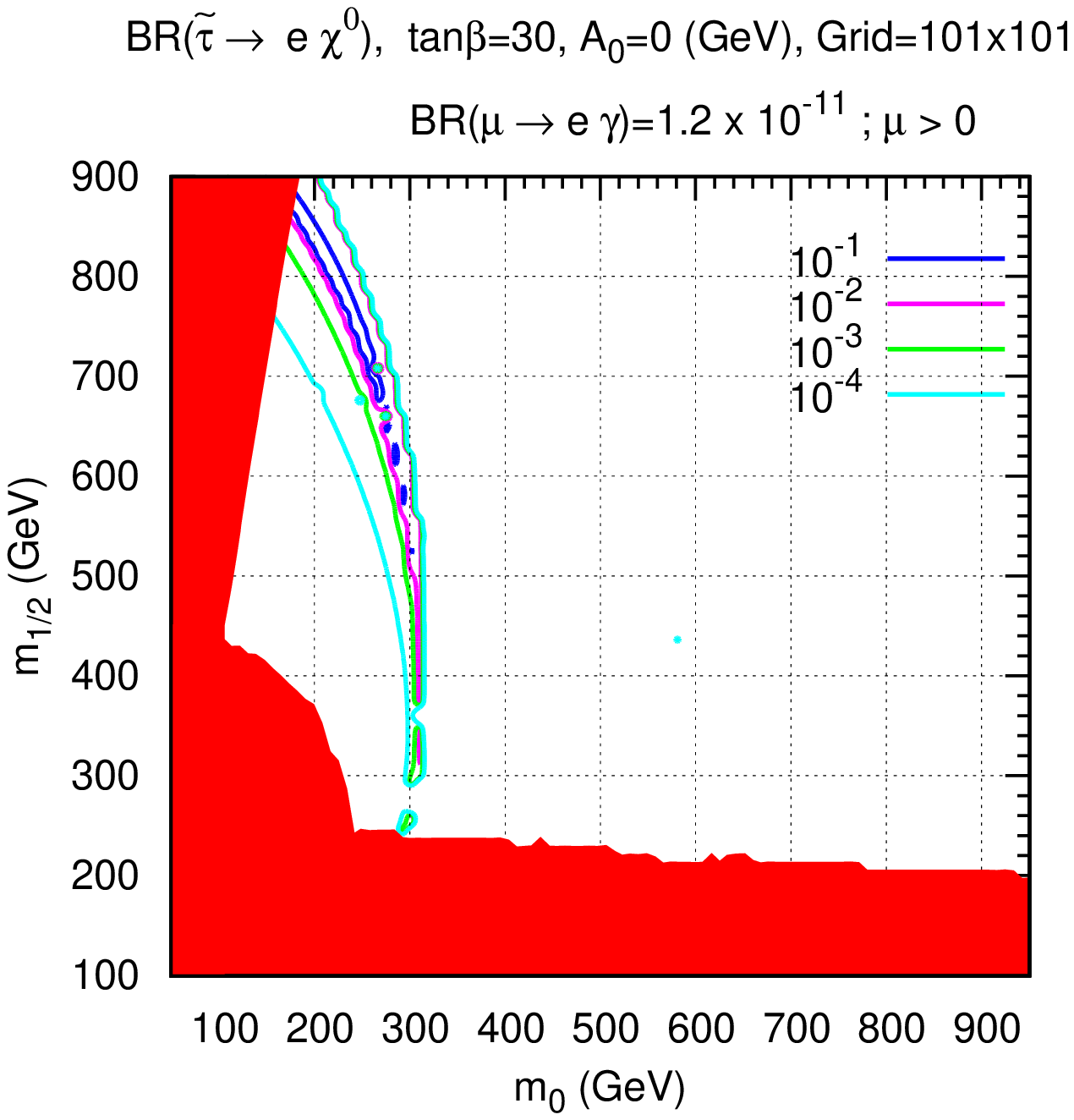}\\\hline
\hspace{0.30cm}%$\vspace{-2cm}
\begin{rotate}{90}{{\small$\tan\beta=10,\  A_0=-300\textrm{ GeV},\  \mu>0$}}\end{rotate} &
  \includegraphics[width=0.35\textwidth]{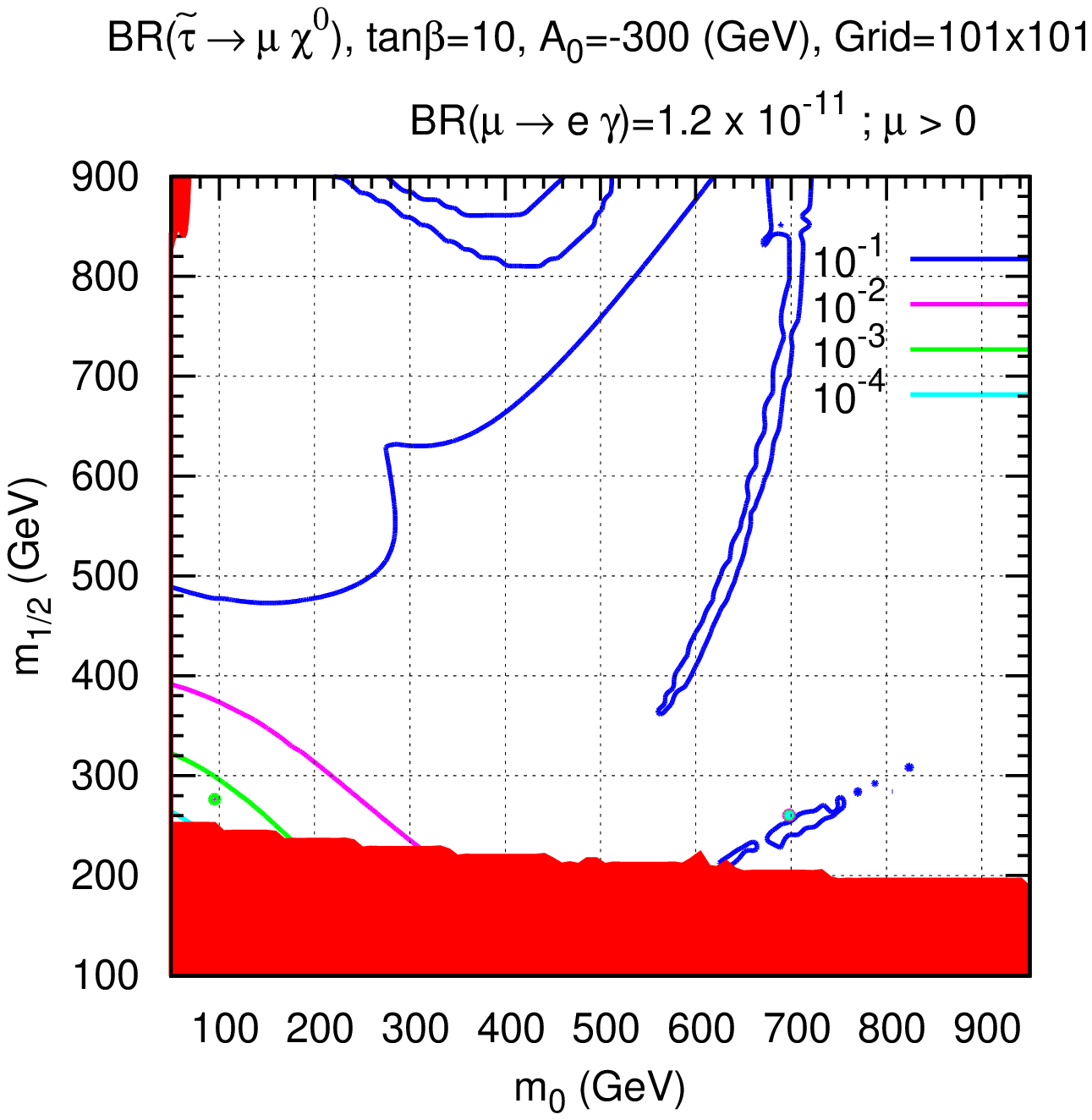}&
  \includegraphics[width=0.35\textwidth]{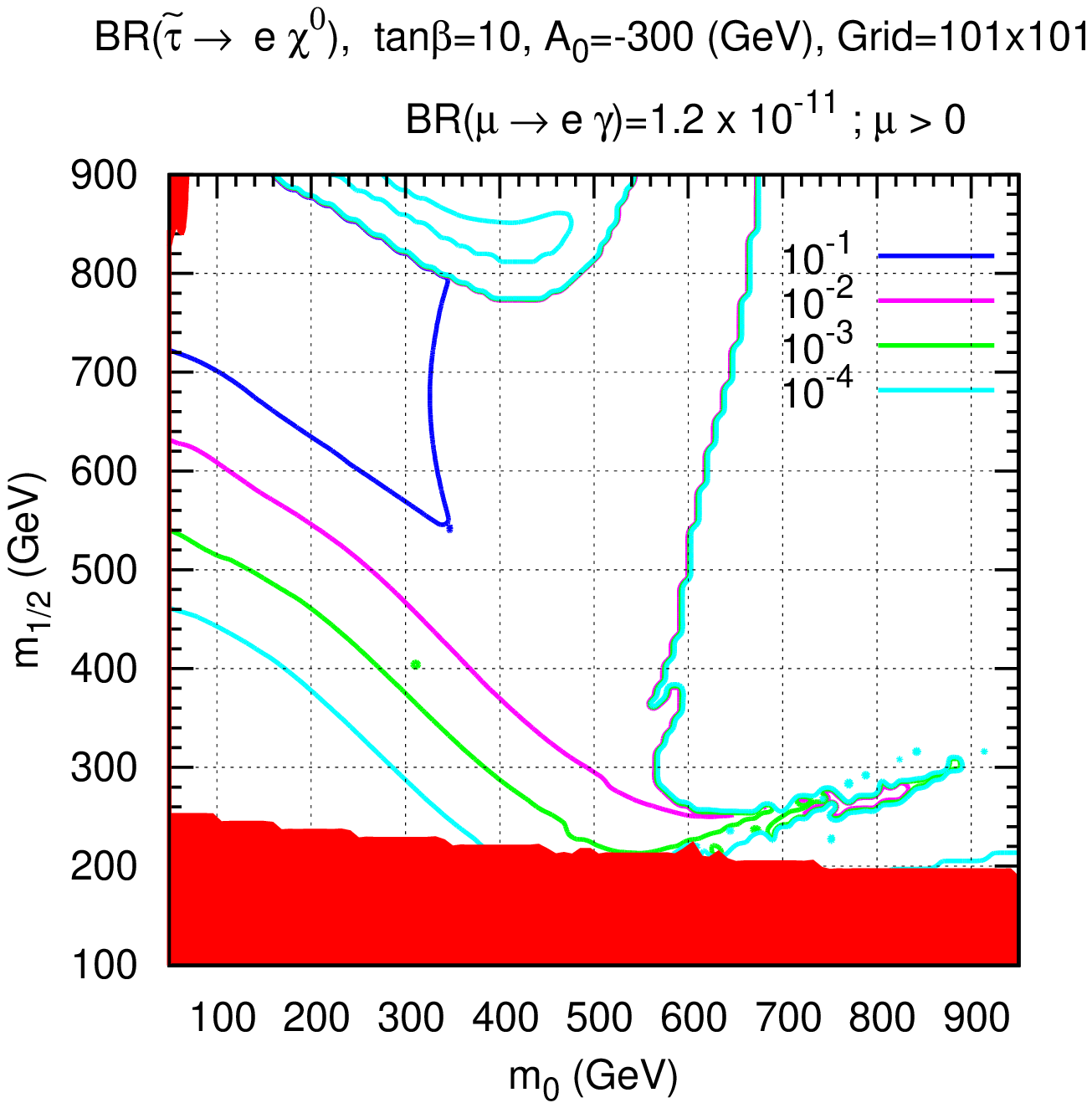}\\\hline
\end{tabular}
  \caption{BR(${\tilde\tau}_2 \to \mu\, \chi^0_1$) (left column) and
    BR(${\tilde\tau}_2 \to e\, \chi^0_1$) (right column) in the $m_0-m_{1/2}$ plane, 
    for the type-II seesaw. We impose BR($\mu\to e\, \gamma) \le 1.2\cdot
    10^{-11}$, and we consider different choice of mSugra parameters:
    $\mu>0$, $\tan\beta=10$ and $A_0=0$ GeV (first row);
     $\mu>0$, $\tan\beta=30$ and $A_0=0$ GeV (second row);
     $\mu>0$, $\tan\beta=10$ and $A_0=-300$ GeV. (third row)} 
  \label{fig:StauLFV-table-II}
\end{figure}

%%%%%%%%%%%%%%%%%%%%%%%%%%%%%%%%%%%%%%%%%%%%%%%%%%%%%%%%%%%%%%%%%%%%%%
In order to simplify our numerical analysis in type-II seesaw, we have considered only the case in which $\lambda_1=2\times 10^{-2}$ and $\lambda_2=0.5$. We have explicitly checked that $\lambda_1$ plays no role in LFV. On the other hand, larger values of $\lambda_2$ generally increase the LFV signal, as BR($\mu\to e\,\gamma$) decreases for larger $\lambda_2$.
Figure~\ref{fig:StauLFV-table-II} shows the same as figure~\ref{fig:StauLFV-table-I}, but for type-II seesaw.
Note that the region where ${\tilde\tau}_2$ LFV BR's is smaller and a deformed with respect the one for type-I seesaw. The reason for this is that the addition of non-gauge-singlet states in type-II seesaw increases the dependence on the renormalization scale of the neutrino Yukawa coupling and also affects the SUSY spectrum (and thus the region where BR($\mu\to e\,\gamma$) is strongly suppressed). 

\begin{figure}[htb]
  \centering
  \includegraphics[width=0.75\textwidth]{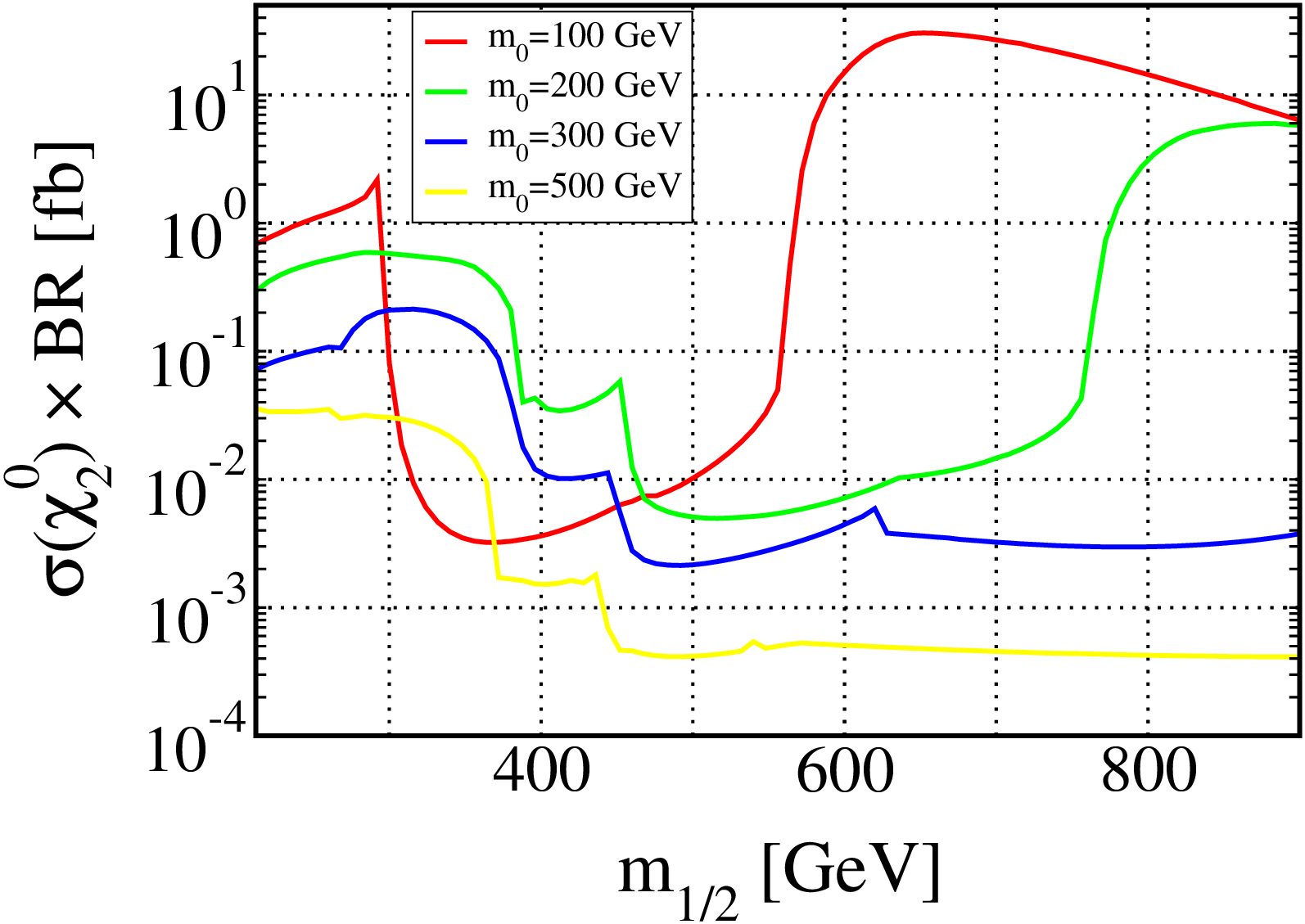}
  \caption{Production cross section (at leading order) of $\chi^0_2$ times BR($\chi^0_2\to\chi^0_1\,\mu\,\tau$)
%    of $\chi^0_2$ going to $\mu$-$\tau$ opposite-sign dilepton signal 
    versus $m_{1/2}$ for $m_0=100$~GeV (red),
    200~GeV (green), 300~GeV (blue) and 500~GeV (yellow). We take a standard choice of
    parameters: $\mu>0$, $\tan\beta=10$ and $A_0=0$ GeV, for type-II seesaw, for
    BR($\mu\to e +\gamma) \le 1.2\cdot 10^{-11}$.} 
  \label{fig:ProdXBR-II}
\end{figure}

In figure~\ref{fig:ProdXBR-II} it is plotted the same as in figure~\ref{fig:ProdXBR-I}, but for type-II seesaw.
Assuming a luminosity of ${\cal L} = 100 fb^{-1}$, there can be a maximum number of events of the order of $10^3$ for $m_0\sim 100$ GeV and $m_{1/2}\sim[600,\, 800]$ GeV.
%%%%%%%%%%%%%%%%%%%%%%%%%%%%%%%%%%%%%%%%%%%%%%%%%%%%%%%%%%%%%%%%%%%%%%%%%%%%%%%%%%%%%%%%%%%%%%%%%%%%%%%%%%%%%%%%%%%%%%%%%%%%%%%%%%%%%%%%%%%%%%%%%%%%%%%%%%%%%%%%%%%%%%
\section{Conclusions}
%%%%%%%%%%%%%%%%%%%%%%%%%%%%%%%%%%%%%%%%%%%%%%%%%%%%%%%%%%%%%%%%%%%%%%%%%%%%%%%%%%%%%%%%%%%%%%%%%%%%%%%%%%%%%%%%%%%%%%%%%%%%%%%%%%%%%%%%%%%%%%%%%%%%%%%%%%%%%%%%%%%%%%
We have shown that neutrino parameters can be indirectly tested at the LHC by measuring the \emph{ratio} of ${\tilde\tau}_2$ LFV BR's in type-I and II SUSY seesaw models if mSugra is assumed. 
We have performed a numerical analysis of the absolute values of ${\tilde\tau}_2$ LFV BR's and we have estimated the maximum number of events that can occur at the LHC. We have shown that there exist regions of the mSugra parameter space where the number of events of the opposite-sign dilepton signal $\chi^0_2\to\chi^0_1\,\mu\,\tau$ can be as much as of the order of $10^3$. 
%%%%%%%%%%%%%%%%%%%%%%%%%%%%%%%%%%%%%%%%%%%%%%%%%%%%%%%%%%%%%%%%%%%%%%%%%%%%%%%%%%%%%%%%%%%%%%%%%%%%%%%%%%%%%%%%%%%%%%%%%%%%%%%%%%%%%%%%%%%%%%%%%%%%%%%%%%%%%%%%%%%%%%
\ack
%%%%%%%%%%%%%%%%%%%%%%%%%%%%%%%%%%%%%%%%%%%%%%%%%%%%%%%%%%%%%%%%%%%%%%%%%%%%%%%%%%%%%%%%%%%%%%%%%%%%%%%%%%%%%%%%%%%%%%%%%%%%%%%%%%%%%%%%%%%%%%%%%%%%%%%%%%%%%%%%%%%%%%
The author wishes to thank his collaborators J~N~Esteves, M~Hirsch, W~Porod, J~C~Romao and J~W~F~Valle.
The author is supported by {\it Funda\c c\~ao para a Ci\^encia e a Tecnologia} under the grant SFRH/BPD/30450/2006.
This work has been partially supported by {\it Funda\c c\~ao para a Ci\^encia e a Tecnologia} through the projects CERN/FP/83503/2008, POCI/81919/2007 and CFTP-FCT UNIT 777 (which are partially funded through POCTI (FEDER)), and by the Marie Curie RTN's MRTN-CT-2006-035505.
%%%%%%%%%%%%%%%%%%%%%%%%%%%%%%%%%%%%%%%%%%%%%%%%%%%%%%%%%%%%%%%%%%%%%%%%%%%%%%%%%%%%%%%%%%%%%%%%%%%%%%%%%%%%%%%%%%%%%%%%%%%%%%%%%%%%%%%%%%%%%%%%%%%%%%%%%%%%%%%%%%%%%%
\section*{References}
\bibliography{iopart-num}
%%%%%%%%%%%%%%%%%%%%%%%%%%%%%%%%%%%%%%%%%%%%%%%%%%%%%%%%%%%%%%%%%%%%%%%%%%%%%%%%%%%%%%%%%%%%%%%%%%%%%%%%%%%%%%%%%%%%%%%%%%%%%%%%%%%%%%%%%%%%%%%%%%%%%%%%%%%%%%%%%%%%%%
\end{document}